# Optimal Coordinated Planning Amongst Self-Interested Agents with Private State

**Ruggiero Cavallo**
Div. Eng. & Applied Sci.
Harvard University
cavallo@eecs.harvard.edu

**David C. Parkes**
Div. Eng. & Applied Sci.
Harvard University
parkes@eecs.harvard.edu

**Satinder Singh**
Computer Sci. and Eng.
University of Michigan
baveja@umich.edu

## Abstract

Consider a multi-agent system in a dynamic and uncertain environment. Each agent's local decision problem is modeled as a Markov decision process (MDP) and agents must coordinate on a joint action in each period, which provides a reward to each agent and causes local state transitions. A social planner knows the model of every agent's MDP and wants to implement the optimal joint policy, but agents are self-interested and have private local state. We provide an incentive-compatible mechanism for eliciting state information that achieves the optimal joint plan in a Markov perfect equilibrium of the induced stochastic game. In the special case in which local problems are Markov chains and agents compete to take a single action in each period, we leverage Gittins allocation indices to provide an efficient factored algorithm and distribute computation of the optimal policy among the agents. Distributed, optimal coordinated learning in a multi-agent variant of the multi-armed bandit problem is obtained as a special case.

## 1 Introduction

Consider a multi-agent system in a dynamic and uncertain environment in which there is a situation of *strategic interdependence*. Specifically, suppose that individual agents are self-interested in that each cares only about maximizing its own payoff, and that agent behaviors are interdependent in that some actions are incompatible with others (imagine, e.g., competition for a shared resource that is required for any agent to act). In this paper we address the *social planning* problem for such settings.

Suppose that a central planner has knowledge of each agent's local world model and initial state, and has been given the authority to make decisions on the agents' behalf. To do optimal planning a significant distributed problem remains, as the current state of each agent is private to that agent, and the planner must somehow gain access to the *true* state information to execute a system-optimal plan.

For example, consider the problem of operating a taxi cab dispatch. Each cab driver is an agent, whose state can be considered the location of his taxi at any given point in time. Drivers receive "reward" (payment) when they have a fare, the magnitude of which is well-known given the route request made by a client to the dispatcher. The dispatcher's task is to do optimal planning: assign cabs to clients in a way that maximizes total profit. However, cab drivers may have incentive to misreport their location in order to be allocated more fares. Here, as in a vast array of other multi-agent problems, a coordination mechanism is required to disarm individuals' self-interest in order to reach socially-optimal outcomes.

In our solution, the planner solicits claims from agents about their current state each period, proposes and enforces the optimal joint action based on agent reports, and then collects specific payments in order to bring truthful reporting (which enables system-optimal planning) into a Markov perfect equilibrium (MPE) of the induced stochastic game. The MPE property means that truthful reporting is the optimal strategy for an agent at any point in time regardless of past group behavior, given that other agents also play the equilibrium going forward, and given common knowledge that the planner has the correct models for all agents. In equilibrium each agent plays a simple stationary strategy (truthful reporting), and thus equilibrium strategies have the Markov property.

In the special case in which local problems are Markov chains and agents compete to take a single action in each period, we leverage the Gittins index to provide an efficient factored algorithm. We obtain optimal coordinated learning in a multi-agent variant of the multi-armed bandit problem as a special case.

### 1.1 Related work

Broadly, this work is situated within the recent literature on applying mechanism design to dynamic environments. For example: *online mechanism design* (OMD) [Friedman and Parkes, 2003; Parkes and Singh, 2003] considers an MDP setting in which there is a global state, agents dynamically arrive and depart, and each agent has private information about its reward function. Truth-revealing in equilibrium can be achieved by asking an agent to make a single claim about its reward function upon arrival. In coordinated planning agents are persistent and need to provide continual information about their private state to the mechanism; the OMD solution does not apply because it does not *keep* providing incentives for an agent to remain truthful. Another example is *online auctions* [Lavi and Nisan, 2000; Hajiaghayi et al., 2004], which provides a stronger form of truthfulness (namely *dominant-strategy* equilibrium) than in our work but at the expense of a far simpler model. We are not aware of any work (based on Groves mechanisms or otherwise) on mechanism design in our setting.

Fudenberg & Tirole [2000] provide a good reference for game-theoretic solution concepts in dynamic games, including Markov perfect equilibrium (MPE). Prior work in economics appears to be limited to (complete-information) *Nash* implementation in dynamic environments [Jackson and Palfrey, 2001], and work on revenue-optimal sequential auctions [Gallien, 2005, e.g.]. In this work we briefly discuss optimal Bayesian learning as an application of our methods: the *learning algorithm* we propose is achieved in an MPE. Previously, Brafman and Tennenholtz [2004] defined the related notion of an *efficient learning equilibrium* (ELE), for two agents in a non-Bayesian setting, without subgame perfection, and with agents able to observe each others' rewards. Learning as a method to *solve* games has been considered both in the game theory literature [Fudenberg and Kreps, 1993, e.g.] and in computer science [Littman, 2001, e.g.]. However, only Kalai and Lehrer [1993] (for a repeated stage game) have considered "rational learning" in which equilibrium of learning *algorithms* is required.

## 2 Preliminaries

We consider a scenario in which a set of agents $I = \{1, \ldots, n\}$ interacts with the world, where each agent is motivated only to maximize its own individual reward over time. Each agent has a representation of the world as it pertains to itself, modeled as a Markov Decision Process (MDP). That is, each agent $i$'s world model consists of: a state space $S_i$, a set of potential behaviors or actions $A_i$, a non-deterministic transition function $\tau_i : S_i \times A_i \times S_i \to [0, 1]$ that defines a probability distribution over successor states for every possible action taken in every possible state, and a reward function $r_i : S_i \times A_i \to \mathcal{R}_{\geq 0}$ that determines the reward received when a particular action is taken in a particular state. Each agent's value for receiving reward decays exponentially by a factor $\gamma$ every time-step into the future, so a reward of $x$ received $t$ steps in the future is valued at $\gamma^t x$.

In a multi-agent setting actions can be interdependent and certain joint actions may be infeasible; i.e., an action taken by one agent may constrain the set of actions available to another agent. For instance, consider a single shared resource required for any action, so that only a single agent can act in any given time-step.

We propose a *coordination mechanism*, in which a central planner selects a joint action, causing a state transition for some or all of the agents. Until the final section on learning, we will assume the center has perfect knowledge of each MDP. However, the local state of each agent's MDP is private. The task of the center is to implement the optimal (feasible) joint policy $\pi^*$ that maximizes total system-value across agents, i.e.,

$$\pi^*(s) \in \arg\max_{\pi \in \Pi_f} \mathbb{E}\Big[\sum_{t=t_0}^{\infty} \sum_{i \in I} \gamma^{t-t_0} r_i(s_i^t, \pi(s^t)) \Big| \pi, s^{t_0} = s\Big],$$

$\forall s \in S$, at all times $t_0 \geq 0$. Here, $\pi$ is a joint policy that at time $t$ maps joint state $s^t$ to a joint action $a^t$, where $s^t \in S : S_1 \times \ldots \times S_n$, $a^t \in A : A_1 \times \ldots \times A_n$, and $\Pi_f$ is the set of all feasible joint policies (i.e., those that respect constraints on the joint actions). We include a *null action* as part of every $A_i$, to model the fact that some agent(s) acting in certain ways may preclude other agents from acting at all. We can also define joint transition function $\tau : S \times A \times S \to [0, 1]$.

The planner can compute (e.g., through value iteration) the optimal value function $V^* : S \to \mathcal{R}_{\geq 0}$, which is the system value of the optimal joint policy $\pi^*$, and from which $\pi^*$ can easily be read off. But, in following the policy the planner must be aware of the present joint state, while the agents, if queried, may have incentive to misreport this information in order to maximize their personal reward.

The tools of *mechanism design* can be used to yield an implementation that achieves the optimal joint policy in equilibrium, despite agent self-interest.

A coordination mechanism $\Gamma = <\pi, T, \mu>$ specifies a joint execution policy $\pi$, a transfer policy $T$ that defines payments *from the planner to the agents* in every joint state, and a joint message space $\mu = \mu_1 \times \ldots \mu_n$ (usually the joint state space $S$). The mechanism can enforce action decisions and transfers in each period, which are made based on the agents' reports. Transfers, $T(s)$, can be negative, implying "charges" rather than payments. Each agent makes claims about in-formation related to its private state in each period and has a *strategy*, $\sigma_i$, which maps the *history* $h$ (the agent's actions, local state trajectory, and transfers received) and current state $s_i^t$ to a message $\sigma_i(h, s_i^t) \in \mu_i$.

The appropriate solution concept for this environment is the Markov Perfect Equilibrium (MPE) [Maskin and Tirole, 2001]. Here is an *informal* definition:

**Definition 1. (Markov Perfect Equilibrium)** *A strategy profile $(\sigma_1^*, \ldots, \sigma_n^*)$ is an MPE if: (a) no agent can improve its expected utility by deviating in any state reachable either on or off the equilibrium path, given the other agents' strategies and the agent's belief about the other agents' private state and local MDP models; (b) the agent updates its beliefs according to Bayes' rule where possible[1]; (c) every strategy is conditioned only on the local state of the agent.*

The equilibrium is *perfect* since the immediate message sent by an agent coupled with its future strategy maximizes its expected utility in all states (in and out of equilibrium play), and *Markov* because each agent's strategy ignores history and simply chooses a message based only on the current state. In general, one also needs Bayesian updating because expectation is w.r.t. Markov dynamics in an agent's MDP and an agent's beliefs about the local state—and model—of other agents.[2]

A *system-optimal* mechanism implements the optimal joint policy. In order to make execution of the mechanism feasible, certain other properties are desirable. For instance, in a *truthful* mechanism agents report truthful information in equilibrium.[3] A mechanism is termed *ex post individual rational (IR)* if it is guaranteed that no agent will ever be worse off from having participated; when this holds only in expectation the mechanism is *ex ante IR*. Likewise, if net transfers from the mechanism to the agents are guaranteed to be no greater than zero, the mechanism is *ex post budget-balanced*; it is *ex ante budget-balanced* when this holds in expectation. A mechanism is *strongly budget-balanced* when net payments from the planner to agents are exactly zero.

## 3 Coordinated Planning: The General Setting

In this section we describe mechanisms for coordinated planning in a general MDP environment. The first mechanism we examine is a sequential version of the basic Groves mechanism [Groves, 1973], which pays each agent a quantity equal to the total reward stream that other agents in the system receive (yielding the "global reward" for each agent if reports are truthful).

---

**Mechanism 1. (Sequential-Groves)**
- *The planner computes the optimal joint policy $\pi^*$.*
- *At every time-step $t$:*
  1. *Each agent $i$ reports to the planner a claim about its current state $\hat{s}_i^t$.*
  2. *The planner implements the joint action $a^t = \pi^*(\hat{s}^t)$.*
  3. *The planner pays each agent $i$ a transfer:*
  $$T_i(\hat{s}^t) = \sum_{j \in I \setminus \{i\}} r_j(\hat{s}_j^t, a_j^t)$$

---

Realize that the transfer made by the planner is received as "reward" by each agent at the same time as the intrinsic reward for the joint action is received.

**Theorem 1.** *The Sequential-Groves mechanism is truthful, system-optimal, and ex post IR in Markov Perfect Equilibrium when agents have a common discount factor.*

*Proof.* Assume all agents except some agent $i$ follow a truthful strategy with $\sigma_i^*(h, s_i) = s_i$ in all states. Let $\nu_i^t(s^t, M)$ denote $i$'s expected payoff from time $t$ to infinity, given that the current joint state is $s^t$, given models $M = (M_1, \ldots, M_n)$ of the MDPs for each agent, and given that agent $i$ is also truthful. Let $\pi^*$ denote the policy of the planner. At *any* time $t_0$, for *any* state $s$ and for *any* models $M$, agent $i$'s expected reward when playing its truthful strategy is:

$$\nu_i^{t_0}(s, M) = \mathbb{E}_M \Big[ \sum_{t=t_0}^{\infty} \Big\{ \gamma^{t-t_0} r_i(s_i^t, \pi^*(s^t)) + \sum_{j \in I \setminus \{i\}} \gamma^{t-t_0} r_j(s_j^t, \pi^*(s^t)) \Big\} \Big| \pi^*, s^{t_0} = s \Big]$$

$$= \mathbb{E}_M \Big[ \sum_{t=t_0}^{\infty} \sum_{j \in I} \gamma^{t-t_0} r_j(s_j^t, \pi^*(s^t)) \Big| \pi^*, s^{t_0} = s \Big] \quad (1)$$

For any $M$, agent $i$ knows that the planner has the correct models $M$ and thus that $\pi^*$ is optimal. Moreover, in any stage of the game, irrespective of whether play was equilibrium or not until now, in equilibrium going forward the planner will know the true state because agents are truthful in equilibrium. By the principle of one deviation take any state $s$ and consider whether agent $i$ can do better with a misreport now and truthful reporting in the future. Since the planner's policy is actually constructed to solve (1), agent $i$'s payoff will be maximized by reporting its true state.

---
[1] Generally, while on the equilibrium path.

[2] But we will achieve a strong form of MPE: each agent will maximize its expected utility whatever the current state, and whatever the model, of other agents. This will make the requirement on Bayesian updating irrelevant.

[3] This is without loss of generality via an appeal to an appropriate generalization of the revelation principle to this setting.

Since the planner implements the optimal policy and agents are truthful, system-optimality is achieved. Ex post IR holds trivially, since agent payoffs always consist of some non-negative reward extracted from the world, plus a non-negative transfer payment $T_i$. □

Desirable properties notwithstanding, this mechanism is clearly unsatisfactory due to its budgetary properties—in expectation it requires $(n-1)\cdot\mathbb{E}[V^*(s^0)]$ to implement, as the value each agent receives each time-step is paid to the $n-1$ other agents.

Truthfulness is obtained by completely aligning the goals of every agent in the system, via the transfer payments *Sequential-Groves* prescribes. Importantly, this remains true even when we impose a separate "charge" on agents to try to eliminate the deficit the payments cause, as long as each agent's charge is *completely beyond its control*.

The Vickrey-Clarke-Groves (VCG) mechanism (see, e.g., [Jackson, 2000]) modifies the basic Groves mechanism by charging each agent $i$ a quantity equal to the value that other agents in the system *could* have received if $i$ were not present; we call this the *VCG charge*. In VCG each agent's transfer payment is the Groves payment minus this charge, and will be equivalent to its marginal contribution to social welfare.

Complications arise in implementing VCG in a sequential environment, as care must be taken to ensure that nothing an agent reports in one time-step is used in computing its VCG charge in any successive time-steps. However, given that the planner knows the agent MDPs and the initial joint state $s^0$, it can compute the expected value $V^*_{-i}(s^0) = \mathbb{E}_M\left[\sum_{t=0}^{\infty}\sum_{j\in I\setminus\{i\}}\gamma^t r_j(s_j^t, \pi^*(s^t))\,\Big|\,\pi^*\right]$, that agents except $i$ would achieve from the optimal joint policy $\pi^*$, and use this to determine VCG-like charges.[4] We propose the following:

---

**Mechanism 2. (Sequential-VCG)** *Identical to the Sequential-Groves mechanism, except at every time t, transfer payments are computed as follows:*

$$T_i(\hat{s}^t) = \sum_{j\in I\setminus\{i\}} r_j(\hat{s}_j^t, a_j^t) - (1-\gamma)V^*_{-i}(s^0)$$

---

**Theorem 2.** *The Sequential-VCG mechanism is truthful, system-optimal, ex ante IR, and ex ante strong budget-balanced in Markov Perfect Equilibrium when agents have a common discount factor.*

---

[4]Note that this is not the value from the optimal joint policy in the marginal problem without $i$, but rather the other agents' value for the actual policy, so it's not completely analogous to the standard VCG charge.

*Proof.* The argument for truthfulness (and subsequently for system-optimality) holds from truthfulness of *Sequential-Groves*, plus the fact that each agent's VCG charge is independent of anything it reports. In equilibrium, the expected total payoff to any agent $i$ at time 0 if it believes that the initial state is $s$ is now as follows (for MDP models $M$):

$$\nu_i^0(s, M) = \mathbb{E}_M\left[\sum_{t=0}^{\infty}\gamma^t r_i(s_i^t, \pi^*(\hat{s}^t))\,\Big|\,\pi^*, s^0=s\right] + \quad (2)$$

$$\mathbb{E}_M\left[\sum_{t=0}^{\infty}\sum_{j\in I\setminus\{i\}}\gamma^t r_j(\hat{s}_j^t, \pi^*(\hat{s}^t))\,\Big|\,\pi^*, s^0=s\right] - \quad (3)$$

$$\sum_{t=0}^{\infty}\gamma^t(1-\gamma)V^*_{-i}(s) \quad (4)$$

$$= \mathbb{E}\left[\sum_{t=0}^{\infty}\gamma^t r_i(s_i^t, \pi^*(\hat{s}^t))\,\Big|\,\pi^*, s^0=s\right] \quad (5)$$

The Groves payment (3) and the VCG charge (4) cancel out in expectation, and so in expectation each agent receives exactly the reward it extracts from the world under the system-optimal policy. Likewise, since the expected payment from the planner to each agent is zero, in expectation the mechanism is strongly budget-balanced. □

### 3.1 Stronger IR in episodic worlds

While the *Sequential-VCG* mechanism succeeds in many respects, it still achieves only ex ante IR: payoffs could end up being negative, depending on what actual path of execution is realized.

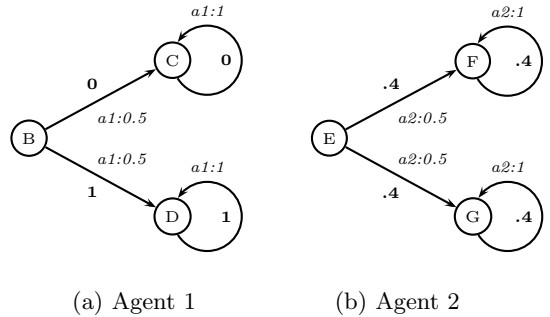

(a) Agent 1      (b) Agent 2

Figure 1: MDPs for a 2-agent world, each with 3 states. If agent 1 executes action *a1* initially, depending on the state-transition, in all future time-steps he will get reward either 0 or 1. Actions and transition probabilities are italicized.

Consider the example in Figure 1. When the system is in joint state $\{B, E\}$, the expected (intrinsic) reward each period to agent 1 going forward is: $0.5\cdot 0 + 0.5\cdot 1 = 0.5$. Agent 2's expected payoff (including transfers) each period is $0.4 + 0.5 - 0.5 = 0.4$ (its reward, plus Groves payment, minus a term representing the per-period equivalent of the expected value to agent 1, i.e. $(1 - \gamma)[0.5(\frac{1}{1-\gamma}) + (0.5)0] = 0.5)$. However, if the state transition $B \to C$ is realized, then agent 2's total payoff per-period each time-step from time 2 on would be $0.4 + 0 - 0.5 = -0.1$, since the realized reward per period to agent 1 is zero. Note that despite this potential for violation of (ex post) IR, reporting truthful state information (including from this state) still maximizes expected payoff. But since it's only ex ante IR, agent 2 would sometimes like to decommit from the mechanism.

To obtain a stronger IR property, we would like to be able to use the *actual* realized joint state at every time-step to compute VCG charges. However, any such charge would not be agent-independent and could produce incentives for agents to misreport their state. Though the above example is too simple to demonstrate this, if circumstances were such that agent 2 could affect which actions were selected in agent 1's MDP through its own state reports, it could potentially gain from misreporting in order to lead the system to a state where agent 1 has no possibility of receiving reward and thus decreasing 2's VCG charge.

Some domains will allow a stronger form of IR. The *Sequential-VCG* mechanism will be ex ante IR from *any* time at which the agent MDPs are in a joint state, known to the planner, that is independent of anything that's ever been reported. Consider worlds in which a certain known-state is guaranteed to be visited periodically, for instance worlds that start in the same state every morning. In this case we can provide ex ante IR periodically, rather than just once—agents will willingly "sign up" for the mechanism repeatedly, regardless of the interim execution, every time the known-state is visited. Similar examples can be provided if periodic monitoring is possible, so that the joint state is known for sure from time to time.

## 4 Coordinated Planning: Markov Chains

In this paper we are specifically exploring coordination mechanisms for achieving *optimal* planning. However, in the general MDP setting computing an optimal policy is typically intractable, as the joint state transition matrices are of size exponential in the number of agents. Nonetheless, there are instances in which optimal planning is possible; we now consider one such special case.

Suppose that each agent's world model is a Markov chain, and at every time-step a single action can be taken. So each agent knows exactly how to behave given the state of its world, but the planning problem remains as only one agent can act per time period. For example, consider the problem of choosing computational tasks for execution on a server. Say there is a set of jobs to be processed, each associated with a known client. Assume the planner has a model of how execution of any job will go, while the actual execution realized is non-deterministic. There is one action for each job in this environment ("process job"), so the agent MDPs are actually Markov chains. Since jobs that are not being processed don't change state, there will be just a single Markov chain activation and local state transition per period.

Formally, such environments can be modeled by specifying that every agent has action space $A_i = \{a_i, a_{\text{null}}\}$, and at every time-step the planner chooses a joint action for which all but one component is $a_{\text{null}}$. For convenience, we write $\pi(s) = i$ to indicate that policy $\pi$ chooses to activate $i$'s Markov chain

Remarkably, Gittins [1974; 1989] showed that in this environment optimal planning can be done in time linear in the number of Markov chains. Specifically, he provided a way of computing an "index" for each Markov chain (given its current state) independent of the state of all others, and showed that the policy that always chooses to activate the Markov chain with the highest index is optimal.

**Theorem 3.** [Gittins and Jones, 1974; Gittins, 1989] *Given Markov chains $M_1, \ldots, M_n$ in states $(s_1, \ldots, s_n)$ respectively, there exist independent functions $G_1(M_1, s_1), \ldots, G_n(M_n, s_n)$ such that the optimal policy $\pi^*(s) = \arg\max_i G_i(M_i, s_i)$.*

In this environment computation of an optimal policy neatly decomposes into $n$ independent sub-problems—each computing the index for one Markov chain—thus avoiding an exponential blow-up in the complexity of optimal planning. Several methods for computing Gittins indices have been proposed, including one by Katehakis and Veinott [1987] in which a special type of two-action, $k$-state MDP is formulated for every state in a $k$-state Markov chain, the optimal value of which corresponds to the Gittins index. Thus all Gittins indices for $n$ distinct $k$-state Markov chains can be computed by solving $n \cdot k$ two-action, $k$-state MDPs.

A potential issue in applying Mechanism 2 is in computing $V_{-i}^*(s^0)$. An index-based algorithm provides a way to simulate the trajectory of an optimal policy but does not provide direct value-function information. Simulation is sufficient for our purposes, though. We can compute an unbiased estimate of $V_{-i}^*(s^0)$ by simulating a policy $\pi$ from initial state $s^0$. This will yield a sample execution path $X_\pi$. We define $r(X, t)$ to be the system reward yielded at time $t$ in sample path $X$. We propose the following coordination mechanism for finite-state Markov chain environments, in which $m$ different sample trajectories $\{X_{\pi^*}^1, \ldots, X_{\pi^*}^m\}$ are maintained (for some $m \geq 1$).

**Mechanism 3. (Sequential-Gittins-VCG)**

- *The planner computes $G_i(M_i, s_i), \forall i \in I, s_i \in S_i$*
- *At every time-step $t$:*

  1. *Each agent $i$ reports to the planner a claim about its current state $\hat{s}_i^t$.*
  2. *The planner activates Markov chain:*
  $$i^* \in \arg\max_{i \in I}\{G_i(M_i, \hat{s}_i^t)\}$$
  *and simulates the next action in each of the $m$ sample trajectories.*
  3. *The planner pays each agent $i$ a transfer:*
  $$T_i(\hat{s}^t) = \begin{cases} -\sum_{k=1}^{m} \frac{r(X_{\pi^*}^k, t)}{m} & \text{for } i^* \\ r(\hat{s}_{i^*}^t) - \sum_{k=1}^{m} \frac{r(X_{\pi^*}^k, t)}{m} & \text{for } j \in I \setminus \{i^*\} \end{cases}$$

**Theorem 4.** *The Sequential-Gittins-VCG mechanism is truthful, system-optimal, ex ante IR, and ex ante strong budget-balanced in Markov Perfect Equilibrium when agents have a common discount factor. The mechanism can be implemented in time linear in the number of agents.*

The proof of the theorem follows from that of Theorems 2 and 3 once one considers that the sample estimates are unbiased. The mechanism's formal properties hold regardless of the number of samples $m$ we use (as long as $m \geq 1$); however, as we choose a higher $m$ the variance in the estimate will decrease.

### 4.1 A distributed implementation

We now consider a scenario in which the agents have computational capacity and can contribute to solving the planning problem. In *Sequential-Gittins-VCG* the computation is centralized (all done by the planner); we now seek a mechanism in which the planner's role is primarily organizational, with the agents themselves performing the computation. Gittins' factored method leads to an immediate distributed computational architecture—each agent will be responsible for computing Gittins index information on its Markov chain. Now we must bring both this computation and truthful reporting of state information into MPE.

The agent-independence of the VCG charge presents a new problem: agent $i$ cannot be involved in the computation to determine its charge. So for each $i$ we consider the reward that would be received under $\pi^{-i}$: *the optimal policy in the marginal world without $i$*. This can be determined by simply observing the highest Gittins index among agents other than $i$, and then *simulating* what would have happened if we activated that chain. Thus to approximate $\pi^{-i}$ we need to compute a distinct set of sample trajectories $\{X_{\pi^{-i}}^1, ..., X_{\pi^{-i}}^m\}$ for each agent $i$. In total, the planner maintains $n \cdot m$ sample trajectories. We propose the following:

**Mechanism 4. (Distributed-Gittins-VCG)**

- *Each agent $i$ computes and reports a claim to the planner about Gittins indices $\hat{G}_i(M_i, s_i), \forall s_i \in S_i$*
- *At every time-step $t$:*

  1. *Each agent $i$ reports to the planner a claim about its current state $\hat{s}_i^t$.*
  2. *The planner activates Markov chain:*
  $$i^* \in \arg\max_{i \in I}\{\hat{G}_i(M_i, \hat{s}_i^t)\}$$
  *and simulates the next action in each of the $n \cdot m$ sample trajectories.*
  3. *The planner pays each agent $i$ a transfer:*
  $$T_i(\hat{s}^t) = \begin{cases} -\sum_{k=1}^{m} \frac{r(X_{\pi^{-i^*}}^k, t)}{m} & \text{for } i^* \\ r(\hat{s}_{i^*}^t) - \sum_{k=1}^{m} \frac{r(X_{\pi^{-j}}^k, t)}{m} & \text{for } j \in I \setminus \{i^*\} \end{cases}$$

**Theorem 5.** *The Distributed-Gittins-VCG mechanism is truthful, system-optimal, ex ante IR, and ex ante weak budget-balanced in Markov Perfect Equilibrium when agents have a common discount factor.*

*Proof.* The proof of truthfulness in MPE (and subsequently system-optimality) is essentially analogous to that of Theorem 2. Each agent's report now consists of computed Gittins indices in addition to current state information. Since the report of Gittins indices by agent $i$ has no effect on its VCG charge (due to the distinct sample trajectories per agent), the interests of each agent are completely aligned with that of the entire system. If agents other than $i$ report truthfully then agent $i$ maximizes its expected value by reporting correct Gittins indices and its true state in every period. The mechanism remains ex ante IR. The total expected VCG charge over time for an agent $i$, given initial state $s$, for any Markov chain models $M$ is:

$$\mathbb{E}_M\Big[\sum_{t=0}^{\infty} \sum_{j \in I \setminus \{i\}} \gamma^t r_j(s_j^t, \pi_{-i}^*(s_j^t)) \Big| \pi_{-i}^*, s^0 = s\Big] \quad (6)$$

$$\leq \mathbb{E}_M\Big[\sum_{t=0}^{\infty} \sum_{j \in I \setminus \{i\}} \gamma^t r_j(s_j^t, \pi^*(s_j^t)) \Big| \pi^*, s^0 = s\Big] + \quad (7)$$

$$\mathbb{E}_M\Big[\sum_{t=0}^{\infty} \gamma^t r_i(s_i^t, \pi^*(s_i^t)) \Big| \pi^*, s^0 = s\Big] \quad (8)$$

The sum of (7) and (8) is agent $i$'s expected payoff under the policy the planner implements. The inequality holds by virtue of the fact that $\pi^*$ is an optimal policy.

The Groves payments (7) made to each agent $i$ will in expectation be less than or equal the VCG charges extracted from that agent (6), since the latter is derived from a policy that is *optimal* for the system without $i$. Since this holds for every agent, the expected net sum of payments from the planner to the agents is non-positive, yielding ex ante weak budget-balance. □

Each agent can compute its Gittins indices in parallel, and the only added time burden on the planner as the number of agents grows is in comparing the indices to determine the maximal index in each period and in simulating trajectories for the VCG charge. Including agents in the computation of Gittins indices precludes the strong budget-balance of the previous mechanisms; thus we see an inherent tradeoff between distributing computation and certain economic properties.

### 4.2 Optimal Coordinated Learning in Multi-Armed Bandit Problems

One of the most compelling aspects of Gittins' result is that it can be used for (computationally) efficient and optimal Bayesian learning in the multi-armed bandit problem (MABP). Here is the standard multi-armed bandit story: a single agent is faced with $n$ arms to pull, each with an uncertain (stationary) reward distribution. The agent has prior information and seeks to maximize its (subjective) expected discounted reward. Each arm can be modeled as a Markov chain: the states are now information states and not physical states, the initial state is defined by prior information, the action is to pull an arm, and the state transition is defined by Bayes' rule. An optimal policy for a MABP performs *optimal* (Bayesian) exploration vs. exploitation and maximizes the expected discounted reward.

We provide a multi-agent variation of the MABP: there is a distinct agent that receives reward for each arm, and there is a planner. Each agent has a (privately held) prior on the reward its arm yields and receives private observations of the reward it receives. Each agent is self-interested and would prefer that its arm is always pulled. The planner, on the other hand, is interested in optimal joint learning. This is an instance of coordinated planning with Markov chains and private state, the solution to which brings optimal joint learning into a game-theoretic equilibrium.

Figure 2 shows the Markov chain representation of a Bernoulli process that has yielded reward 0 twice and 1 once in three activations so far, displaying two subsequent time-steps. State transition $\tau_i(s_i, s_i')$ for agent $i$ represents the probability of transitioning from information state $s_i$ to $s_i'$ when $i$'s arm is chosen, given the current belief about the arm as represented in state $s_i$.

In an infinite horizon environment, the Markov chain for each arm will technically be infinite in size. How-ever, since reward $t$ steps in the future is discounted by $\gamma^t$, we can effectively ignore states far enough removed when computing each index.[5] Nonetheless, since the Gittins index for *any* state could still potentially be required, the infinite state space precludes any mechanism that elicits all Gittins indices in the initial step.

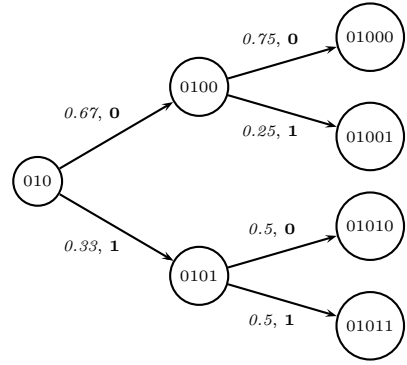

Figure 2: Markov chain representation of two activation steps of a Bernoulli bandit, with 3 previous observations (010). Transition probabilities are in italics.

Instead, we elicit index reports from agents online, for both the actual execution path and the simulated marginal world trajectories. In addition, we no longer assume the initial models are common knowledge and instead elicit them from the agents in the first step. For this we need a concise language, such as a representation of priors as sample observations (e.g., 01011).

---

**Mechanism 5. (Learning-Gittins-VCG)**

- *Each agent reports its prior beliefs to the planner.*
- *At every time-step $t$:*

  1. *Each agent $i$ communicates to the planner a claim about its current state $\hat{s}_i^t$ and corresponding Gittins index $\hat{G}_i(M_i, \hat{s}_i^t)$.*
  2. *The planner activates Markov chain:*
     $$i^* \in \arg\max_{i \in I} \{\hat{G}_i(M_i, \hat{s}_i^t)\}$$
  3. *For each agent $i$, the planner updates sample execution paths $\{X_{\pi_{-i}}^1, ..., X_{\pi_{-i}}^m\}$, asking agents in $I \setminus \{i\}$ to report Gittins indices for states on the frontier of any sample path.*
  4. *The planner pays each agent $i$ a transfer:*
     $$T_i(\hat{s}^t) = \begin{cases} -\sum_{k=1}^{m} \dfrac{r(X_{\pi_{-i^*}}^k, t)}{m} & \text{for } i^* \\ r(\hat{s}_{i^*}^t) - \sum_{k=1}^{m} \dfrac{r(X_{\pi_{-j}}^k, t)}{m} & \text{for } j \in I \setminus \{i^*\} \end{cases}$$

---

[5]This can be formalized by bounding the reward in any state with some $R_{\max}$ and modeling agents as $\epsilon$-indifferent so that they will play an $\epsilon$-MPE.

**Theorem 6.** *The Learning-Gittins-VCG mechanism implements optimal Bayesian learning in the multi-agent MABP in a Markov Perfect Equilibrium when agents have a common discount factor. The planner needs space $O(n)$ and runs in time $O(n^2)$ per iteration.*

*Proof Sketch.* Optimal Bayesian learning is achieved by implementing the optimal joint policy (recall that states are now information states). Recognizing this, the proof is then largely analogous to that of Theorem 5. Each agent will choose to report its correct model (via its prior) and report truthful state information and correct Gittins indices. (Actually, it will be weakly indifferent about these reports since the planner's decisions about policy only use its index reports—see below.) Moreover, agent $i$ will report correct Gittins indices in each marginal world trajectory (for $\pi^{-j}$, $j \neq i$) since these reports do not affect its reward. Fixing $m$, the planner needs $O(n)$ space to track the actual and simulated trajectories and $O(n)$ time to make the optimal decision on *each* trajectory in each period (i.e., finding the maximal index). □

*Learning-Gittins-VCG* is also ex ante IR and ex ante weak budget-balanced for the same reasons as *Distributed-Gittins-VCG*.[6] The online computation in this mechanism leads to cases of *weak indifference* in agent reporting of models, state, and Gittins indices. Thus, some remarks regarding the robustness of the proposed learning mechanism are in order: (a) For reporting the model, if an agent $j \neq i$ reports an incorrect model but reports Gittins indices for the same (incorrect) model then MPE is retained; (b) For reporting the current state, the planner could threaten random checks of consistency between state and Gittins index (with a fine), which is sufficient to remove this weak indifference; (c) For reporting the Gittins indices on the sample trajectory paths the effect of deviation would retain ex ante IR but jeopardize budget-balance (since the marginal policy would be suboptimal and thus the VCG charge less in expectation).

## 5 Conclusion

In this paper we provided coordination mechanisms for a setting in which self-interested agents have MDP models of their worlds (known to the planner) and private state information, where the goal is optimal joint planning. The assumption that MDP models are known is actually not essential as shown in the final application to optimal multi-agent learning. Computing the optimal policy in the general setting remains an issue. In the context of Markov chains we leveraged the Gittins index to achieve computational efficiency and to distribute the computation among the agents; exploring mechanisms that employ distributed policy computation in other settings could be a fruitful line of future work.

---

[6]Realize though that the "strength" of the IR provision depends on the accuracy of initial priors: when initial knowledge about each action is limited the approximation of the VCG charge we execute could be little better than a random guess at the true VCG charge we'd compute if we could use all information reported by the agents throughout execution of the mechanism. We think about this as *very* ex ante IR, although formally it remains ex ante IR!